\documentclass[showpacs,twocolumn,superscriptaddress,nofootinbib]{revtex4}

\usepackage{doi}
\usepackage{hyperref}
\hypersetup{
  colorlinks=true,        
  linkcolor=blue,         
  citecolor=cyan,        
}

\usepackage{graphicx}
\usepackage{dcolumn}
\usepackage{bm}
\usepackage{color}
\usepackage{xcolor}

\usepackage{amsmath}
\usepackage{amssymb}


\begin{document}

\title{Weak cosmic censorship conjecture in the pure Lovelock gravity }

\author{Sanjar Shaymatov}
\email{sanjar@astrin.uz}

\affiliation{Institute for Theoretical Physics and Cosmology, Zheijiang University of Technology, Hangzhou 310023, China}
\affiliation{Akfa University,  Milliy Bog Street 264, Tashkent 111221, Uzbekistan}
\affiliation{Ulugh Beg Astronomical Institute, Astronomy St. 33, Tashkent 100052, Uzbekistan}
\affiliation{National University of Uzbekistan, Tashkent 100174, Uzbekistan}
\affiliation{Tashkent State Technical University, Tashkent 100095, Uzbekistan}

\author{Naresh Dadhich}
\email{nkd@iucaa.in}

\affiliation{Inter University Centre for Astronomy \&
Astrophysics, Post Bag 4, Pune 411007, India }

\date{\today}
\begin{abstract}
It is well known that a rotating black hole in four dimension could be overspun by linear order test particle accretion which however always gets overturned when non-linear perturbations are included. It turns out that in the Einstein gravity, repulsion due to rotation dominates over attraction due to mass in dimensions, $D>5$, and consequently black hole cannot be overspun even for linear order accretion. For the pure Lovelock rotating black hole, this dimensional threshold is $D>4N+1$ where $N$ is degree of  single $N$th order term in the Lovelock polynomial in the action. Thus the pure Lovelock rotating black holes always obey the weak cosmic censorship conjecture (WCCC) in all dimensions greater than $4N+1$. Since overall gravity being repulsive beyond this dimensional threshold, how is rotating black hole then formed there?
\end{abstract}
 \maketitle

\section{Introduction}
\label{introduction}

Since cosmic censorship hypothesis
\cite{Penrose69} has remained unproven yet, testing the weak cosmic censorship conjecture (WCCC) and the physical possibility of destroying
the event horizon of black holes with infalling test particles or fields have become an object of active scientific research. However, in general
relativity (GR), a naked singularity is still an unanswered question. Thus, the theoretical existence of naked singularity is important because their existence would lead to observation of gravitational collapse in diverging gravitational field regime. The formation of naked singularity has been discussed in various collapse models~\cite{Christodoulou86,Joshi93,Joshi00,Goswami06,Harada02,Stuchlik12a,Vieira14,Stuchlik14,Giacomazzo-Rezzolla11,Joshi15}.
Black holes have now come to the center stage by the detection of stellar mass black hole mergers by the LIGO and
Virgo scientific collaborations~\cite{Abbott16a,Abbott16b}. It is expected that these observations would lead to probing the hidden properties of black holes through gravitational waves produced by mergers or collisions of these objects.

The question that whether a black hole could be overspun to destroy event horizon and thereby laying bare central singularity was first addressed by Wald \cite{Wald74b} who showed that an extremal black hole can never be overspun; i.e. extremal horizon cannot be destroyed by test particle accretion. Much later it was also shown that a non-extremal black hole cannot be converted into extremal one by imploding it by test particles of suitable parameters \cite{Dadhich97}. The interest in this question was revived when Jacobson and Sotirio (JS)\cite{Jacobson10} argued that though extremality could neither be reached nor destroyed but it could perhaps be jumped over. Beginning with a near extremal state, they had shown that test particles of appropriate parameters could be thrown in a discrete manner such that black hole could overspin without passing through the extremal state. This was however preceded by consideration of overcharging a charged Reissner-Nordstr\"{o}m black hole by Hubeny \cite{Hubeny99}. Later, it was also shown by Saa and Santarelli \cite{Saa11} that it would be possible
to destroy a Kerr-Newman black hole horizon with a charged particle accretion. Of late the question of turning black hole into naked singularity by test particle accretion has become an active field of interest. An  extensive discussion of the possibility of destroying event horizon by bombarding near extremal black hole by test particles in various
background geometry can be found in \cite[see,
e.g.][]{Matsas07,Shaymatov15,Bouhmadi-Lopez10,Rocha14,Jana18,Song18,Duztas18,Duztas-Jamil18b,Duztas-Jamil20,Yang20a,Yang20b,Gwak20,Ayyesha22PDU}. In all these works,  particles are assumed to follow the test particle trajectories, and the back reaction effects were not included. In fact impossibility of
destroying black hole horizon relies on radiation reaction and
self-force effects for which infalling particles could be
unable to impinge onto black hole~\cite{Barausse10,Zimmerman13,Rocha11,Isoyama11,
Colleoni15a,Colleoni15b,Li13}. It turns out that a near extremal Kerr black hole can be turned into a Kerr-Newman naked singularity, however it is interesting to note that introduction of a test magnetic field acts as a cosmic censor restoring WCCC~\cite{Shaymatov15}. Inclusion of back reaction
effects have been investigated for possibility of destroying
regular black hole horizon~\cite{Li13}.

A large amount of effort has also been devoted to the study of WCCC for black hole surrounded by a complex scalar test fields~\cite[see,
e.g.][]{Toth12,Duztas13,Duztas14,Semiz15,Gwak19JCAP}, a rotating anti-de Sitter black
holes~\cite{Gwak16JCAP,Natario16,Natario20,Gwak21JCAP}, magnetized black holes~\cite{Siahaan16,Shaymatov19b}, and black holes in Einstein-Gauss-Bonnet gravity~\cite{Ghosh19}. The WCCC was also probed by various tests~\cite{Mishra19} in relation to the laws of black hole dynamics~\cite{Bardeen73b}. Recently Chakraborty et al. \cite{Chakraborty17} worked out a nice discriminator between black hole and naked singularity by considering spin precession of spinning test particle in the vicinity of black hole and naked singularity spacetimes.

There is the famous Myers-Perry solution~\cite{Myers-Perry86} for a higher dimensional rotating black hole. The case of five and six dimensional black holes has been thoroughly analysed for WCCC in~\cite{Shaymatov19a,Shaymatov20a}. In higher dimensions, the relative dominance between attraction due to mass and repulsion due to rotation is expected to play a critical role. It turns out that repulsion would override attraction for large $r$ in $D>5$. It is shown by explicit calculation that six dimensional black hole cannot be overspun even for linear order test particle accretion~\cite{Shaymatov20a}. This is because of resultant gravity being repulsive, test particles would be unable to reach the horizon.   Since dominance of repulsion over attraction would be in all dimensions $>5$, therefore black hole in all dimensions higher than five cannot be overspun~\cite{Shaymatov21a} and thereby the WCCC would be always obeyed. 

There is another related issue of non-existence of bound orbits around black holes in higher dimensions in the Einstein gravity \cite{Dadhich13}. The cause for this is whether the centrifugal force is able to balance the gravitational attraction. While in the case of rotating black hole in higher dimensions, repulsive gravitational effect of rotation begins overriding over attraction due to mass. It turns out the resultant force for higher dimensional rotating black hole would always be repulsive at infinity~\cite{Dadhich22b}. That is also reflected in the fact that black hole cannot be overspun even under linear accretion process in $D > 5$. However, the situation gets overturned when one considers pure Gauss-Bonnet (GB)/Lovelock gravity. That is in the dimension window $2N + 2 \leq D \leq 4N$ dimensions, there do exist bound and stable circular orbits only around pure GB rotating black holes~\cite{Dadhich22a}.

Even though black hole may be overspun at linear order accretion, yet when non-linear perturbations are included in accordance with the Sorce-Wald analysis~\cite{Sorce-Wald17}, the result is always overturned~\cite{An18,Gwak18a,Ge18,Shaymatov21:EPJC,Ning19,Yan-Li19,Shaymatov21d,Jiang20plb,Shaymatov19c}. Since in dimensions greater than five, there is no possibility of overspinning even at the linear order, there is no question of its being overturned at non-linear order. In this way WCCC is indeed always obeyed for a rotating black hole in all dimensions greater than five.

It has been strongly argued~\cite{Dadhich16a} that pure Lovelock equation, which involves the only one $N$th order term in the Lovelock action without sum over the lower orders, is the proper gravitational equation in higher dimensions. For probing WCCC in higher dimensions, we should therefore study possibility of overspinning pure Lovelock rotating black holes. Though there exists the exact solution for a static pure Lovelock black hole~\cite{Dadhich11GB}, however there exists no exact solution for a pure Lovelock rotating black hole.  A pure Gauss-Bonnet rotating black hole metric was constructed~\cite{Dadhich13b} by using the static black hole solution on the same lines~\cite{Dadhich13GB} as the Kerr metric was done from the Schwarzschild solution. This metric has all the desirable properties of a rotating black hole and hence could be taken as a good effective metric. It has been employed by authors in studying its energetics and other properties~\cite{Abdujabbarov15a,Dadhich-Ghosh13}. Here we would go a step further in constructing a pure Lovelock metric as an analogue of the higher dimensional Myers-Perry rotating black hole solution~\cite{Myers-Perry86}. We argue as follows. The spacetime would have the same symmetry as that of the Myers-Perry metric, the only thing that should change is  gravitational potential due to mass which we would replace by that of pure Lovelock potential of a static black hole. Like the pure GB case it would not be an exact solution of the pure Lovelock vacuum equation, yet it would have all the features of a higher dimensional rotating black hole giving a good effective description. We shall employ this metric to examine WCCC for pure Lovelock rotating black hole.

In Sec.~\ref{Sec:Lovelock} we briefly describe rotating pure Lovelock black hole metric and its properties which is followed by the properties of pure GB rotating black hole in six dimension in Sec~\ref{Sec:GB}. In Sec.~\ref{Sec:Perturbation} we present variational identities for Einstein-Maxwell-Gauss-Bonnet (EMGB) theory to derive linear and non-linear perturbation inequalities. In Sec.~\ref{Sec:GN}
we probe the WCCC for pure GB rotating black holes for linear and non-linear test particle accretion by applying gedanken experiments.  We end with discussion and conclusions in Sec.~\ref{Sec:Conclusions}. Throughout we use a system of units in which $G=c=1$.

\section{Pure Lovelock rotating black hole}\label{Sec:Lovelock}

The Myers-Perry higher dimensional rotating black hole in dimension $D$ \cite{Myers-Perry86,Dadhich13GB} is described by the metric, %
\begin{eqnarray}\label{D}
ds^2&=&-dt^2+r^2d\beta^2 + (r^2+a^2_{n})\left(d\mu_{n}^2+\mu_{n}^2d\phi^2_{n}\right)\nonumber\\&+&\frac{\mu r}{\Pi F}\left(dt +a_{n}\mu_{n}^2d\phi_{i}\right) +\frac{\Pi F}{\Delta}dr^2\, ,
\end{eqnarray}
with
\begin{eqnarray}
F &=& 1-\frac{a_{n}^2\mu_{n}^2} {r^2+a_{n}^2}\, , \nonumber\\
\Pi &=&(r^2+a_1^2)...(r^2+a_n^2) \, ,
\end{eqnarray}
and
\begin{eqnarray}\label{Eq:delta1}
\Delta = \frac{\left(r^2+a^2\right)...\left(r^2+a_{n}^2\right)}{r^{2n-2}} -2\mu r^{2n+3-D}\, ,
\end{eqnarray}
where $n=[(D-1)/2]$ is the maximum number of rotation parameters in dimension $D$. Here $\mu$ and $a_n$ refer respectively to mass and rotation parameters while $\mu_n$ and $\beta$ are related respectively for even and odd $D$ as follows:
\begin{eqnarray}
\Sigma \mu_n^2 + \beta^2 &=& 1\, ,\\
\Sigma \mu_n^2 &=& 1\, .
\end{eqnarray}

In Ref.~\cite{Dadhich-Ghosh13}, the pure GB effective metric for a rotating black hole was written by replacing potential $\mu/r^{D-3}$ by the pure GB one, $\mu/r^{(D-5)/2}$. We shall accordingly write instead the pure Lovelock potential of arbitrary order $N$, $\mu/r^\alpha$ where $\alpha=(D-2N-1)/N$~\cite{Dadhich11GB} for the effective metric of a pure Lovelock rotating black hole in dimensions $D>2N+1$.\footnote{$N$ is degree of the Lovelock polynomial in action which has the only one $N$th order term for pure Lovelock. Also note that in the critical odd dimension, $D=2N+1$ pure Lovelock gravity is kinematic meaning vacuum solution is trivial with Lovelock Riemann tensor vanishing~\cite{Dadhich12,Dadhich16b}.} In effect this would simply amount to replacing the power of $r$ in the last term of Eq.~(\ref{Eq:delta1}) by ${2n-\alpha}$. That means $\Delta$ for the pure Lovelock rotating black hole reads as follows:

\begin{eqnarray}\label{Eq:delta2}
\Delta = \frac{\left(r^2+a^2\right)...\left(r^2+a_{n}^2\right)}{r^{2n-2}} -2\mu r^{2n-\alpha}\, ,
\end{eqnarray}
where $\alpha=(D-2N-1)/N$. Thus we have obtained an effective metric for the pure Lovelock analogue of the Myers-Perry rotating black hole. As mentioned earlier it is not an exact solution of the pure Lovelock vacuum equation though it satisfies the equation in the leading order, and has all the required properties for a rotating black hole.

The horizon of black hole is located at the real positive roots of $\Delta=0$ which is given by
\begin{eqnarray}\label{Eq:delta3}
(r^2+a_1^2)...(r^2+a_n^2) - 2\mu\,r^{\left(2n-\frac{D-2N-1}{N}\right)} =0\, .
\end{eqnarray}

From the above polynomial equation in $D\geq2N+2$ for any Lovelock order, the necessary condition required for overspinning is that black hole should admit two horizons and for that the following condition must hold:
\begin{eqnarray}\label{Eq:con1}
2N+1<D<2N(n+1)+1\, .
\end{eqnarray}

This condition tells that for $N=1$, the only dimensional window open for overspinning is $4\leq D \leq5$ while for $N=2$, it is $6\leq D \leq9$ for $n=1$ and $6 \leq D \leq 13$ for $n=2$. This means for $N=1$, Einstein gravity a black hole cannot be overspun in all dimensions, $D>5$ while for $N=2$,  pure GB it cannot be overspun in $D>9$ for single and $D>13$ for two rotations. This is because under these bounds, a black hole can have only one horizon and hence the question of its overspinning does not arise. 

In higher dimensions, there is yet another consideration that may have a bearing on this question. For a rotating object, gravitational potential has two components, one attractive due to mass and the other repulsive due to rotation. As dimension increases the former becomes sharper while the leading term for the later would always go as $1/r^2$ and hence the latter could dominate over the former at some dimension threshold.

Let us then write gravitational potential $2\Phi(r)={\Delta}/{r^2}-1$ for black hole in $D\geq2N+2$ dimensions for pure Lovelock of order $N$. By recalling Eq.~(\ref{Eq:delta1}), it would be given by
\begin{eqnarray}\label{Eq:phi}
2\Phi(r)=\frac{(r^2+a^2)...(r^2+a_{n}^2)}{r^{2n}}-\frac{2\mu}{r^{(D-2N-1)/N}}-1\, .
\end{eqnarray}
In particular we write it for pure GB, $N=2$ and $n=2$ as
\begin{eqnarray}
2\Phi_{GB}(r)&=&-\frac{2\mu}{r^{(D-5)/2}}+\frac{a_1^2+a_2^2}{r^2}+\frac{a_1^2a_2^2}{r^4}\, .
\end{eqnarray}
The repulsive component will override when $\alpha=(D-2N-1)/N > 2$; i.e. $D > 4N+1$. That is for $D > 4N+1$, resultant gravitational force turns repulsive and hence the question arises in such a situation could particles with appropriate parameters to overspin black hole reach down to the horizon? The answer turns out to be negative. It has been shown by an explicit calculation for the six dimensional Myers-Perry black hole~\cite{Myers-Perry86} that it cannot be overspun by linear order accretion of test particles. Hence this should be the case whenever resultant gravity is repulsive for large $r$; i.e. in all dimensions $D > 4N+1$. This is the stronger constraint than the earlier one given by Eq.~(\ref{Eq:delta3}).

\begin{figure*}
\centering
  \includegraphics[width=0.45\textwidth]{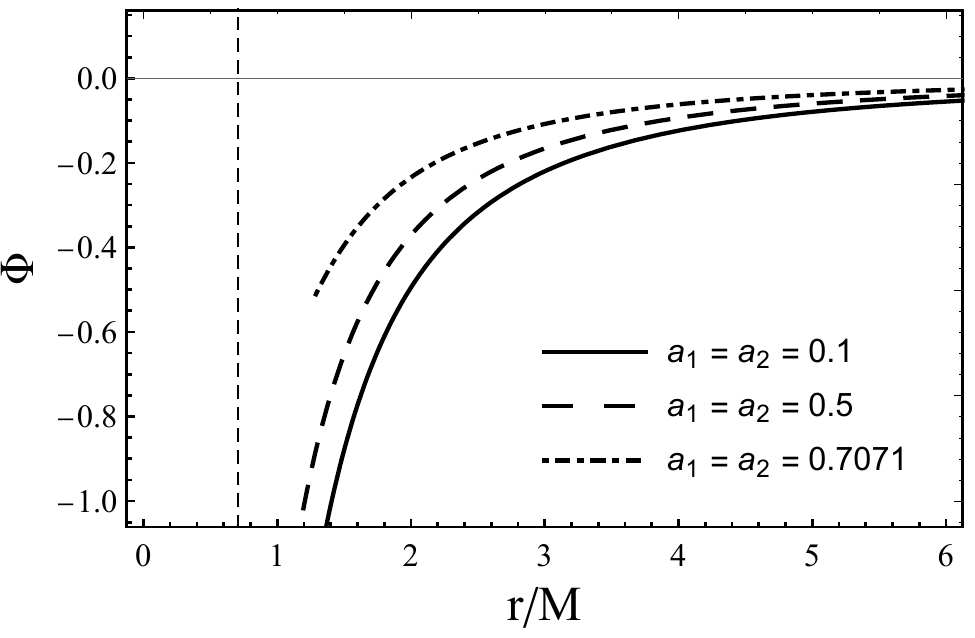}
  \includegraphics[width=0.45\textwidth]{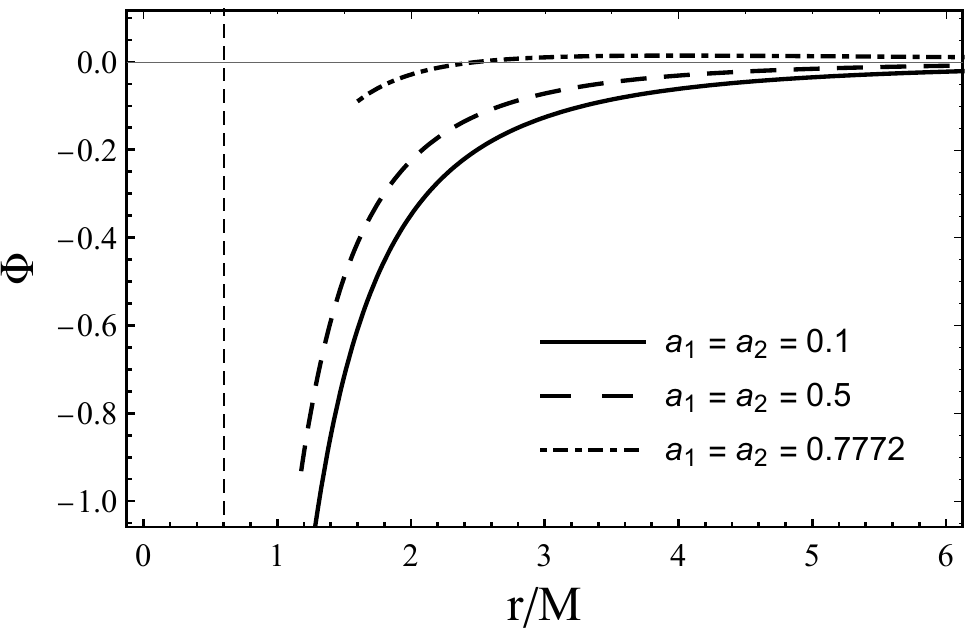}

\caption{\label{fig1} $\Phi(r)$ is plotted for pure GB rotating black hole with $n=2$ rotation parameters in $D=9$ on the left and $D=10$ on the right. The horizon is shown by the vertical dashed line for near extremal black hole for which the plot is indicated by dot-dashed lines. }
\end{figure*}

For pure GB rotating black hole the transition from attraction to repulsion would occur as dimension is crossed from $9$ to $10$. This is shown in Fig.~\ref{fig1}, explicitly for $n=2$. It shows that for near extremal rotation parameters, the potential, $\Phi$ is positive asymptotically. Note that dimensional window available for overspinning of pure Lovelock black hole is given by
\begin{eqnarray}\label{Eq:Dw}
D_w = 4N+2-2N-2 = 2N\, .
\end{eqnarray}
That is for $N=1$, $D_w=2$ (i.e. $D=4,5$) while for $N=2$, $D_w=4$ (i.e. $D=6,...,9$), and so on.

In the next Sec. we shall consider an explicit example of overspinning for pure GB rotating black hole in this allowed  dimensional window.

\section{Pure Gauss-Bonnet rotating black hole with single rotation in six dimension}\label{Sec:GB}

By recalling Eq.~(\ref{Eq:delta2}) we write the following equation for black holes with $n=1$ rotation~\cite{Dadhich-Ghosh13},
\begin{eqnarray}
\Delta &=& r^{2}+a^{2}-2\mu r^{3/2}\, ,\\
\Sigma &=&r^{2}+a^{2}\cos^{2}\theta\ .
\end{eqnarray}

By setting $\mu=\mathcal{M}^{1/2}$ the horizon  radius would be given by $\Delta=0$ which means
\begin{eqnarray}
r^4 - 4\mathcal{M} r^3 + 2a^2r^2 + a^4 = 0\, ,
\end{eqnarray}
and it solves to give
 \begin{eqnarray}\label{lovelock_hor1}
 r_{\pm}&=&{\mathcal{M}}+ \frac{X(\mathcal{M},a)}{\sqrt{6}}\pm
 \left({8\mathcal{M}^2}-\frac{8a^4-2Y^2(\mathcal{M},a)}{3Y(\mathcal{M},a)}\right.\nonumber\\&+&\left. \sqrt{3}\frac{64\mathcal{M}^3-32 a^2 \mathcal{M})}{4\sqrt{2}X(\mathcal{M},a)}-\frac{8a^2}{3}\right)^{1/2}\, ,
 \end{eqnarray}
with
 \begin{eqnarray}
 X^{2}(\mathcal{M},a)&=& 6\mathcal{M}^2-2a^2+ \frac{4a^{4}}{Y(\mathcal{M},a)}+Y(\mathcal{M},a)\, ,
 \nonumber\\
Y^{3}(\mathcal{M},a)&=& {27}\mathcal{M}^2 a^4 -8a^6+3\sqrt{3}\,\mathcal{M}\,
a^4\sqrt{{27}\mathcal{M}^2-16 a^{2}}\nonumber \, ,
 \end{eqnarray}
Note that in the above expression $\mathcal{M}$ and $a$ are respectively related to the black hole mass $M$ and angular momentum $J$ by the following relations:
\begin{eqnarray}\label{Eq:mass-par}
M=\frac{N}{2\pi}\frac{\pi^{(2N+1)/2}}{\Gamma(2N+1)/2}\,\mathcal{M}\, \\
\label{Eq:angular-par}
J=\frac{1}{2\pi}\frac{\pi^{(2N+1)/2}}{\Gamma(2N+1)/2}\,\mathcal{M}\, a\, .
\end{eqnarray}

As can be seen from the above equation the horizons can exist only when $|a|\leq3\sqrt{3}\mathcal{M}/4$ and hence the extremality condition is given by
\begin{eqnarray}\label{Eq:extr-ty}
f &=& 27\mathcal{M}^2-16 a^{2} = 0\,
\end{eqnarray}
and extremal horizon is $r_{+}={9\mathcal{M}}/{4}$.

The area of the event horizon of rotating black hole with single rotation in $D=6$ can be evaluated by setting $dr=dt=0$ and $r=r_{+}$. The horizon metric reads as
\begin{eqnarray}
\label{det} g_{\alpha\beta}=
\left(\begin{array}{cccc} \Sigma & 0 & 0 & 0\\ \\
0 & \frac{\Upsilon(r_{+}) \sin ^2(\theta )}{\Sigma } & 0 & 0\\ \\
0 & 0 & r_{+}^2\cos^2\theta & 0\\ \\
0 & 0 & 0 & r_{+}^2\cos^2\theta\sin^2\psi\\ \\
 \end{array}\right)\, ,
\end{eqnarray}
where $\Upsilon(r_{+})=\left(r_{+}^2+a^2\right)^2$ 
The horizon area is computed as
\begin{eqnarray}\label{Eq:hor-area}
A&=&\int_{\Xi_{4}}\sqrt{det|g_{\alpha\beta}|}d\theta d\phi d\psi d\eta\nonumber\\&=&
2\mathcal{M}^{1/2}r^{7/2}\frac{2\pi^{(2N+1)/2}}{\Gamma(2N+1)/2}\, ,
\end{eqnarray}
which must not decrease in any physical process according to the famous area non-decrease theorem~\cite{Grunau15}.

The angular velocity is given by $\Omega_{+}={a}/{(r_+^2+a^2)}$
for which the Killing field is defined by
\begin{eqnarray}
\label{killingfield}\xi=\xi^\alpha\partial_\alpha=\xi_{(t)}+\Omega_{+}\xi_{(\phi)}\, .
\end{eqnarray}
In terms of the Killing field, the surface gravity is defined by
\begin{eqnarray}\label{sg1}
2k\xi_{\alpha}=\nabla_{\alpha}\left(-\xi_{\beta}\xi^{\beta}\right)|_{r=r_+}\,
,
\end{eqnarray}
or by
\begin{eqnarray}\label{sg2}
k^2=-\frac{1}{2}\Big(\nabla_{\alpha}\xi_{\beta}\Big)\Big(\nabla^{\alpha}\xi^{\beta}\Big)|_{r=r_+}\,
.
\end{eqnarray}
For the metric in question, it evaluates as
\begin{eqnarray}
k&=&\frac{\Upsilon(r)^{-3/2}}{2\Sigma^{1/2}\Delta^{1/2}}\nonumber\\&\times &\left[g_{rr}\Big(g^{rr}F_{1}(r, \theta)\Big)^{2}+g_{\theta\theta}\Big(g^{\theta\theta}F_{2}(r, \theta)\Big)^{2}\right]^{1/2}\, ,
\end{eqnarray}
where we have defined
\begin{eqnarray}
F_{1}(r, \theta)&=&\Upsilon(r)^2g_{tt,r}+4\mathcal{M}(r)a r \Upsilon(r)g_{t\phi,r}\nonumber\\&+&4\mathcal{M}^2(r)a^2r^2g_{\phi\phi,r}\, ,\\
F_{2}(r, \theta)&=&\Upsilon(r)^2g_{tt,\theta}+4\mathcal{M}(r)a r\Upsilon(r)g_{t\phi,\theta}
\nonumber\\&+&4\mathcal{M}^2(r)a^2r^2g_{\phi\phi,\theta}\, ,
\end{eqnarray}
where $\mathcal{M}(r)=\mathcal{M}^{1/2}r^{1/2}$ with $\mathcal{M} $ related to the black hole mass. From the above equations, the surface gravity for six-dimensional rotating pure GB black hole is given by
\begin{eqnarray}\label{Eq:sur-gravity}
k=\frac{2r_{+}^{1/2}-{3}\mathcal{M}^{1/2}}{4\mathcal{M}^{1/2} r_{+}}\, .
\end{eqnarray}

We shall now examine the possibility of overspinning of the black hole.

\section{Varitional identities and perturbation inequalities }\label{Sec:Perturbation}

{A diffeomorphism covariant leads to derive variational identities for given 
manifold in $n$ dimensions.  For that theory one should describe Lagrangian $L$ including $g_{\alpha\beta}$ and other physical fields $\psi$~\cite{Wald94}. The above mentioned Lagrangian's variation is given by 
\begin{eqnarray}\label{Eq:Lag}
\delta L=E \delta\phi+d{\Theta}(\phi,\delta\phi)\, .
\end{eqnarray}
Here we note that $\phi=(g_{\alpha\beta},\psi)$ and $E$, as a parameter of Lagrangian, can be further defined as dynamical fields. Following to that Lagrangian is composed of the fields $\phi$, we can write the equation of motion as ${E}=0$. Note that here ${\Theta}$ is symplectic current $(n-1)$-form and is defined as follows  
\begin{eqnarray}
{\omega}(\phi,\delta_1\phi,\delta_2\phi)=\delta_1{\Theta}(\phi,\delta_2\phi)-\delta_2{\Theta}(\phi,\delta_1\phi)\, ,
\end{eqnarray}
with the corresponding variations $\delta_{1,2}$ .  Meanwhile,  for the vector field $\zeta^{\alpha}$ the Noether current $(n-1)$-form can be written as 
\begin{eqnarray}\label{Eq:Noether}
{J}_\zeta={\Theta}(\phi,{L}_\zeta\phi)-\zeta \cdot{L}\, .
\end{eqnarray}
From the above equation the condition $d{J}_\zeta=0$ satisfies the equation of motion. Following to \cite{Wald95,Sorce-Wald17}, the Noether current takes the following form
\begin{eqnarray}\label{Eq:Noether1}
{ J}_\zeta=d{Q}_\zeta+ {C}_\zeta\, .
\end{eqnarray}
The first term on the right hand side in the above equation, ${Q}_\zeta$, represents the Noether charge, while the term ${C}_\zeta=\zeta^{\alpha}{C}_{\alpha}$ marks the constraint of the theory.  }

Following to (\ref{Eq:Noether}) and (\ref{Eq:Noether1}), we obtain the first order variational identity keeping $\zeta^{\alpha}$ fixed   
\begin{eqnarray}\label{Eq:linear}
\int_{\partial\Xi}\delta {Q}_\zeta-\zeta\cdot{\Theta}(\phi,\delta\phi)&=&\int_{\Xi}{\omega}(\phi,\delta\phi,\mathcal{L}_\zeta\phi)\nonumber\\&-&\int_{\Xi}\zeta\cdot{E}\delta\phi-\int_{\Xi}\delta \mathbf{C}_\zeta\, ,
\end{eqnarray}
where $\Xi$ represents a Cauchy surface. Note that the first term on the right hand side is the variation part and vanishes in the case only $\zeta^{\alpha}$ corresponds to a Killing vector and a symmetry of the field $\phi$ so that $\mathcal{L}_\zeta\phi=0$. Similarly to the first order,  the second order at the same surface can be obtained as   
\begin{eqnarray}\label{Eq:non-linear}
\int_{\partial\Xi}\delta^2 {Q}_\zeta-\zeta\cdot\delta{\Theta}(\phi,\delta\phi)]&=&\int_{\Xi}{\omega}(\phi,\delta\phi,\mathcal{L}_\zeta\delta\phi)\nonumber\\ &-&\int_{\Xi}\zeta\cdot\delta{E}\delta\phi-\int_{\Xi}\delta^2 {C}_\zeta\, .\nonumber\\
\end{eqnarray}

As mentioned above $\zeta^{\alpha}$ is regarded as a Killing vector field, the first order variation, Eq.~(\ref{Eq:linear}), can be further given by 
\begin{eqnarray}\label{Eq:E=0}
\int_{\partial\Xi}\delta {Q}_\chi-\chi\cdot{\Theta}(\phi,\delta\phi)&=&-\int_{\Xi}\delta \mathbf{C}_\chi\, . 
\end{eqnarray}
Here,  $\chi^{\alpha}=\chi_{(t)}^{\alpha}+\Omega_{+}^{(\phi)}\chi_{(\phi)}^{\alpha}$ refers to the Killing vector and in which $\Omega_{+}^{\phi}$ is the horizon angular velocity. The above mentioned the Cauchy surface $\Xi$ is referred to as the bifurcation surface $B$ at one end and spatial infinity at the other. With thin in view, Eq.~(\ref{Eq:E=0}) at the Cauchy surface $\Xi$ can be defined by %
\begin{eqnarray}\label{Eq:B+In}
\int_{\partial\Xi}\delta {Q}_\chi-\chi\cdot{\Theta}(\phi,\delta\phi)&=&\int_{\infty}\delta {Q}_\chi-\chi\cdot{\Theta}(\phi,\delta\phi)\nonumber\\&-&\int_{B}\delta {Q}_\chi-\chi\cdot{\Theta}(\phi,\delta\phi)\, . \nonumber\\
\end{eqnarray}
The first term on the right hand side in the above equation (i.e. gravity part) can be obtained as 
\begin{eqnarray}\label{Eq:Boundary}
\int_{\infty}\delta {Q}_\chi-\chi\cdot{\Theta}(\phi,\delta\phi)&=&\delta M-\Omega_{+}^{(\phi)}\delta
J_{\phi}\, . 
\end{eqnarray}
From the above procedure,  we rewrite Eq.~(\ref{Eq:linear}) for the first order variation 
\begin{eqnarray}\label{Eq:linear-order1}
\delta M-\Omega_{+}^{(\phi)}\delta
J_{\phi}&=&\int_B[\delta {Q}_\chi-\chi\cdot{\Theta}(\phi,\delta\phi)]
-\int_\Xi\delta {C}_\chi\, .\nonumber\\
\end{eqnarray}
The first term in the above equation vanishes at the bifurcation surface $B$ as that of properties of Cauchy surface $\Xi$ (i.e., $\mathcal{L}_\chi\phi=0$ ). 

Similarly, one can also rewrite Eq.~(\ref{Eq:non-linear}) for the second order variation 
\begin{eqnarray}\label{Eq:non-linear1}
\delta^2 M&-&\Omega_{+}^{(\phi)}\delta^2 J_{\phi}\nonumber\\&=&\int_B[\delta^2 {Q}_\chi-\chi\cdot\delta{\Theta}(\phi,\delta\phi)]\nonumber\\&-&\int_\Xi\chi\cdot\delta{E}\delta\phi-\int_\Xi\delta^2 {C}_\chi+\mathcal{E}_\Xi(\phi,\delta\phi)\, . 
\end{eqnarray}
We note that $\mathcal{E}_\Xi(\phi,\delta\phi)$ is canonical energy at the Cauchy surface $\Xi$  and is considered as the second order perturbation to the field $\delta\phi$ .\\

We further consider Einstein-Maxwell-Gauss-Bonnet (EMGB) theory to derive a perturbation inequalities for testing the WCCC in the pure lovelock rotating black hole. We then write the Lagrangian in EMGB theory as follows \cite{Sorce-Wald17,Jiang20plb}: 
\begin{eqnarray}\label{Eq:EM-Lag}
\mathbf{L}=\frac{\mathbf{\epsilon}}{16\pi}\left(R+ \alpha_{GB} \mathcal{L}_{GB} \right)\, ,
\end{eqnarray}
with $\mathbf\epsilon$ and $\alpha_{GB}$ that refer to the volume element for given spacetime metric and GB constant, respectively, and GB term $\mathcal{L}_{GB}$ given by 
\begin{eqnarray}
\mathcal{L}_{GB}=R^2-4R_{\alpha\beta}R^{\alpha\beta}+R_{\alpha\beta\gamma\delta}R^{\alpha\beta\gamma\delta}\, ,
\end{eqnarray}
with the scalar curvature $R$. Here we note that the equation of motion in EMGB gravity is defined by 
\begin{eqnarray}
G_{\alpha\beta}+\alpha_{GB}H_{\alpha\beta}=T_{\alpha\beta}\, ,
\end{eqnarray}
where $H_{\alpha\beta}$ is gven by 
\begin{eqnarray}
H_{\alpha\beta}&=&2R_{\alpha\gamma\delta\tau}R_{\beta}^{\gamma\delta\tau}-4R_{\alpha\gamma\beta\delta}R^{\gamma\delta}-4R_{\alpha\gamma}R_{\beta}^{\gamma}\nonumber\\
&+&2RR_{\alpha\beta}-g_{\alpha\beta}\mathcal{L}_{GB}\, .
\end{eqnarray}
Note that $G_{\alpha\beta}$ and $T_{\alpha\beta}$ refer to the Einstein tensor and the stress-energy tensor of ordinary matter source, respectively.  
For the Lagrangian given in the above, the dynamical field that is defined by the metric function provided that there exist no other fields like electromagnetic field, i.e. $\phi=(g_{\alpha\beta}, \psi=0)$. The equation of motion is then given by  
\begin{eqnarray}
E(\phi)\delta\phi=-\frac{{\epsilon}}{2} T^{\alpha\beta}
\delta g_{\alpha \beta}\, .
\end{eqnarray}
It is then clearly obvious that the symplectic potential consists of only gravity part, and that is given by  
\begin{eqnarray}\label{Eq:sym}
\Theta_{a_2...a_6}\left(\phi,\delta\phi\right)&=&\frac{1}{16\pi}\epsilon_{a_2...a_6 \alpha }\Big(P_{\beta}^{\gamma\alpha\delta} \delta\Gamma^{\beta}_{\gamma\delta}\nonumber\\ &+& \delta g_{\alpha\delta}\triangledown_{\beta}P^{\beta\gamma\alpha\delta}\Big)\, . 
\end{eqnarray}
with 
\begin{eqnarray}
P^{\alpha\beta\gamma\delta}=\frac{1}{2}\left(g^{\alpha\gamma}g^{\beta\delta}-g^{\alpha\delta}g^{\beta\gamma}\right)+\alpha_{GB}\frac{\partial \mathcal{L}_{GB}}{\partial R_{\alpha\beta\gamma\delta}}\, .
\end{eqnarray}
Eq. (\ref{Eq:sym}) leads to obtain the corresponding form of the symplectic current as  
 \begin{eqnarray}
\omega_{a_2...a_6}&=&\frac{1}{4\pi}\Big[\delta_{1}(\epsilon_{a_2...a_6 \alpha } P_{\beta}^{\gamma\alpha\delta}) \delta_{2} \Gamma_{\gamma\delta}^{\beta}\nonumber\\&-&\delta_{2}(\epsilon_{a_2...a_6 \alpha } P_{\beta}^{\gamma\alpha\delta}) \delta_{1} \Gamma_{\gamma\delta}^{\beta}\nonumber\\&+&
\delta_{1}(\epsilon_{a_2...a_6 \alpha } \triangledown_{\beta}P^{\beta\gamma\alpha\delta}) \delta_{2} g_{\gamma\delta}\nonumber\\&-&\delta_{2}(\epsilon_{a_2...a_6 \alpha } \triangledown_{\beta}P^{\beta\gamma\alpha\delta}) \delta_{1} g_{\gamma\delta}\Big]\, ,
\end{eqnarray}
For the Lagrangian in EMGB theory~(\ref{Eq:EM-Lag}), one can define the Noether charge $Q_{\zeta}$ that stems from the Noether current 5-form given by Eq.~(\ref{Eq:Noether1}) with $C_{\zeta}=\zeta \textbf{C}$. Let us then write the Noether charge $Q_{\zeta}$ and the constraint $C_{\zeta}$ of Gauss-Bonnet gravity as  
\begin{eqnarray}
(Q_{\zeta})_{a_2..a_5}&=&-\frac{1}{16\pi}\epsilon_{a_2..a_5 \alpha\beta}\Big(P^{\alpha\beta\gamma\delta}\triangledown_{\gamma}\zeta_{\delta}\nonumber\\&-&2\zeta_{\delta}\triangledown_{\gamma}P^{\alpha\beta\gamma\delta}\Big)
\, ,\nonumber\\
(C_{\alpha})_{a_2..a_6}&=&\epsilon_{a_2..a_6 \beta}\,T_{\alpha}^{\beta}\, . 
\end{eqnarray}

\section{Perturbation inequalities and gedanken experiment to overspin a pure Gauss-Bonnet rotating black hole with single rotation in six dimension}\label{Sec:GN}

{In this section, we derive linear and non-linear order variational identities caused by matter falling into the black hole. That is in-falling matter adds the mass and angular momentum to black hole's mass and angular momentum, respectively. With this way, the black hole is perturbed linearly, i.e., $M+\delta M$ and $Q+\delta Q$, respectively. However, in a realistic process in-falling matter would involve non-linear perturbations. This, in turn, could change the situation completely. Thus, we involve non-liner perturbations in testing the WCCC for pure Gauss-Bonnet rotating black hole in six dimension. For that we utilize new gedanken experiment developed by Sorce and Wald \cite{Sorce-Wald17}. Following this experiment, we shall further consider a one-parameter family of field $\phi(\lambda)$ perturbation, which allows one to write the equation of motion as 
\begin{eqnarray}
G_{\alpha\beta}(\lambda)+\alpha_{GB}H_{\alpha\beta}(\lambda)=T_{\alpha\beta}(\lambda)\, .
\end{eqnarray}
For the above mentioned one-parameter family of field $\phi(0)$, $T_{\alpha\beta}(0)=0$ must be satisfied so that in-falling particles are assumed to cross the black hole's horizon. For one-parameter family of field perturbation we shall define a hypersurface $\Xi=\Xi_{1}\cup H$ involving a region that begins with the bifurcation surface $B$ and extends up to the horizon $H$. Also $\Xi$ extends to the other end where it turns spacelike $\Xi_{1}$, and then attains to the asymptotical flatness at infinity. The pure Gauss-Bonnet rotating black hole geometry is further supposed to be linearly stable under the field $\phi(\lambda)$ perturbation, which  could vanish at the bifurcation surface $B$. For the perturbed field $\phi(\lambda)$ we shall further consider Gaussian null coordinates on the horizon portion $H$, and that is given by  
\begin{eqnarray}\label{Eq:bs}
\int_B\delta \mathbf{Q}_\chi(\lambda)=\frac{k}{8\pi}A_{B}(\lambda)\, ,
\end{eqnarray}
with the bifurcate surface area $A_{B}$ for pure GB rotating black hole.
In spite of this fact, the spacetime geometry turns into perturbed state with $M(\lambda)$ and $J_{\phi}(\lambda)$ at the sufficiently late times as a consequence of the dynamical field caused by in-falling matter. }

{We further obtain the linear order perturbation inequality with the help of Eq.~(\ref{Eq:linear-order1}) by which black hole parameters are perturbed linearly as
\begin{eqnarray}
\label{change_Mass1} \delta M&=&
\int_{H}\epsilon_{a_1...a_6 \alpha}\chi_{(t)}^{\beta} \delta T_{\beta}^{\alpha} \, ,\\
\label{change_angular1} \delta J_{\phi}&=&
-\int_{H}\epsilon_{a_1...a_6 \alpha}\chi_{(\phi)}^{\beta}\delta T_{\beta}^{\alpha}\, , 
\end{eqnarray}
so that it tends another perturbed state. Note that the integration is over surface element on the event horizon $r_{+}$. {As discussed above the first term $\int_B[\delta {Q}_\chi-\chi\cdot{\Theta}(\phi,\delta\phi)]$ on the right hand side of Eq.~(\ref{Eq:linear-order1}) vanishes at the bifurcation surface, and so it yields }
\begin{eqnarray}
\label{Eq:change_Mass2} \delta M&-&\Omega_{+}\delta
J_{\phi}=-\int_\Xi\delta {C}_\chi=
\nonumber\\
&-&\int_{H}\epsilon_{a_1...a_6 \alpha}\left(\chi_{(t)}^{\beta}+\Omega_{+}\chi_{(\phi)}^{\beta}\right)\delta T_{\beta}^{\alpha}\, .
\end{eqnarray}
Here $\chi^{\beta}=\chi_{(t)}^{\beta}+\Omega_{+}\chi_{(\phi)}^{\beta}$ is regarded as null generator of the horizon $r_+$. And so Eq.~(\ref{Eq:change_Mass2}) can be rewritten as
\begin{eqnarray}
\label{Eq:change_Mass3} \delta M&-&\Omega_{+}\delta
J_{\phi}= - \int_{H}\epsilon_{a_1...a_6 \alpha} \chi_{\beta} \delta T^{\beta\alpha}\, .
\end{eqnarray}
As for the volume element on the horizon it takes the form as $\epsilon_{a_1...a_6 \alpha}=-6\,\tilde{\epsilon}_{[a_1...a_6} k_{\alpha ]}$, and it is given by 
\begin{eqnarray}
- \int_{H}\epsilon_{a_1...a_6 \alpha} \chi_{\beta} \delta T^{\beta\alpha}=\int_{H}\tilde{\epsilon}_{a_1...a_6} \chi_{\beta}k_{\alpha} \delta T^{\beta\alpha}\, .
\end{eqnarray}
For the null energy condition that satisfies $\delta T_{\alpha\beta}k^{\alpha}k^{\beta}\geq 0$, we have the form for the linear order perturbation inequality  
\begin{eqnarray}\label{Eq:change_Mass4} 
\delta M-\Omega_{+}\delta
J_{\phi}\geq 0\, .
\end{eqnarray}}

{As for the the non-linear order variational identity we follow the same procedure as the one for the linear order perurbation. Thus, we have 
\begin{eqnarray}\label{Eq:Non1}
\delta^2 M&-&\Omega_{+}\delta^2 J_{\phi}=\int_B[\delta^2 {Q}_\chi-\chi\cdot\delta{\Theta}(\phi,\delta\phi)]\nonumber\\&-&\int_\Xi\chi\cdot\delta{E}\delta\phi-\int_\Xi\delta^2 {C}_\chi+\mathcal{E}_\Xi(\phi,\delta\phi)\nonumber\\ 
&=& \int_B[\delta^2 {Q}_\chi-\chi\cdot\delta{\Theta}(\phi,\delta\phi)]+\mathcal{E}_H(\phi,\delta\phi)\nonumber\\&-&\int_H\chi\cdot\delta{E}\delta\phi\nonumber\\&-&\int_{H}\epsilon_{a_2...a_6\, \alpha}\left(\chi_{(t)}^{\beta}+\Omega_{+}\chi_{(\phi)}^{\beta}\right)\, \delta^2 T_{\beta}^{\alpha}\nonumber\\ &=& \int_B[\delta^2 {Q}_\chi-\chi\cdot\delta{\Theta}(\phi,\delta\phi)]+\mathcal{E}_H(\phi,\delta\phi)\nonumber\\&+& \int_{H}\tilde{\epsilon}_{a_2...a_6} \chi_{\beta}k_{\alpha} \delta^2 T^{\beta\alpha}\, , 
\end{eqnarray}
with the vector $\chi^{\alpha}$ that is tangent to the horizon portion $H$. By imposing the null energy condition $\delta^2 T_{\alpha\beta}k^{\alpha}k^{\beta}\geq0$ we rewrite Eq.~(\ref{Eq:Non1}) in the following form   
\begin{eqnarray}\label{Non2}
\delta^2 M-\Omega_{+}\delta^2 J_{\phi}&=& \int_B[\delta^2 {Q}_\chi-\chi\cdot\delta{\Theta}(\phi,\delta\phi)]\nonumber\\&+&\mathcal{E}_H(\phi,\delta\phi)\, .
\end{eqnarray}
We shall further evaluate Eq.~(\ref{Non2}) using perturbation field $\phi^{GB}(\lambda)$ caused by in-falling matter 
\begin{eqnarray}
\int_B[\delta^2 {Q}_\chi-\chi\cdot\delta{\Theta}(\phi,\delta\phi^{GB})] ~~\mbox{and}~~\mathcal{E}_{H}(\phi,\delta\phi^{GB})\, .
\end{eqnarray}
After the black hole is perturbed, a perturbed stationary state, $\delta\phi^{GB}$ is reached with the parameters 
$M(\lambda)$ and $J_{\phi}(\lambda)$, which involve $\delta M$ and $\delta J_{\phi}$ that are proposed to be the linear order perturbations alluded in Eq.~(\ref{Eq:change_Mass4}). 
The perturbed field $\delta\phi^{GB}$ satisfies $\delta^2 M=\delta^2J_{\phi}=\mathcal{E}_{H}(\phi,\delta\phi^{GB})=0$, and  $\chi^{\alpha}=0$ can be neglected at the bifurcation surface $B$. And so for this family of perturbation non-linear order variational inequality is defined by         
\begin{eqnarray}
\delta^2 M-\Omega_{+}\delta^2 J_{\phi}&=&\int_B[\delta^2 {Q}_\chi-\chi\cdot\delta{\Theta}(\phi,\delta\phi^{GB})]\nonumber\\&\geq & -\frac{k}{8\pi}\delta^2 A^{GB}\, .
\end{eqnarray}
}

{Following the well-accepted version of gedanken experiment as alluded above we should further examine whether a six dimensional pure GB black hole is overspun by linear and non-linear order test particle perturbations. This is what wish to address in the next.} \\

\subsection{Linear order particle accretion through old version of gedanken experiment}
\label{Sec:linear}

 Before applying new version of gedanken experiment as alluded above we first test the old version of gedanken experiment that involves the linear order test particle perturbations. Following the old version, as always we assume that falling in particles parameters are much smaller as compared to the black hole parameters so as to respect the test particle constraints, i.e., $\delta M\ll M $ and
$\delta J\ll J $. Once the falling in particle with these parameters is absorbed by black hole its corresponding parameters would increase and respectively attain to the new perturbed state
with the parameters $M+\delta M$ and $J +\delta J$.
Following the extremality condition given by Eq.~(\ref{Eq:extr-ty}), we rewrite the following inequality
\begin{eqnarray}\label{Eq:Hor1}
 27\mathcal{M}^2 \leq 16 a^{2} \, .
\end{eqnarray}
This clearly shows that black hole horizon no longer exists. Keeping in mind Eqs.~(\ref{Eq:mass-par}) and (\ref{Eq:angular-par}), the above inequality Eq.~(\ref{Eq:Hor1}) defines the minimum threshold value of angular momentum of impinging particle
\begin{eqnarray}\label{eq:min}
{\frac{9\sqrt{3}}{32\pi}} \bigg(M+\delta M\bigg)<\left(\frac{J+\delta J}{M+\delta M}\right)\, .
\end{eqnarray}
We begin with a nearly extremal black hole which is bombarded with particles of appropriate parameters so as to tip it towards overspinning. For a nearly extremal state we write $a={\frac{9\sqrt{3}M}{16\pi}}\left(1-\epsilon^2\right)$ where $\epsilon \ll 1$. Then Eq.~(\ref{eq:min}) leads to
the minimum threshold value which is given by
\begin{eqnarray}\label{Eq:Jmin1}
\delta
J_{min}&=&\frac{9\sqrt{3}}{32\pi}M^{2}\epsilon^2 +\frac{9\sqrt{3}}{16\pi}~M\delta M
+\frac{9\sqrt{3}}{32\pi}\delta M^2\, .
\end{eqnarray}
This is the lower bound on angular momentum required for impinging particle to fall into the black hole.

For particle to fall into the black hole it must reach the horizon and that would define the upper bound given by
\begin{eqnarray}\label{Eq:energy}
\delta M \geq \Omega_{+}\delta J\, ,
\end{eqnarray}
with $\Omega_{+}=a/(r_+^2+a^2)$ being angular velocity of the horizon. Hence, the upper threshold is given by
\begin{equation}\label{Eq:Jmax} \delta
J_{max}=\frac{r^2_{+} + a^2}{a} \delta M \, .
\end{equation}
This is the upper bound on particle's angular momentum, which on substituting for $r_+$ and $a$ we obtain
\begin{eqnarray}\label{Eq:Max}
\delta J_{max}&=&
\bigg(\frac{9\sqrt{3}}{4\pi}
+\frac{27\epsilon}{4\pi}\bigg)\,M\,\delta
E\, .
\end{eqnarray}
For particle to fall into the black hole it must have $\delta J_{max}>\delta J_{min}$ such that when it impinges on black hole it tends to overspin it. The angular momentum window for overspinning $\Delta J = \delta J_{max} - \delta J_{min}$ is then given by
{\begin{eqnarray} \label{Eq:Delta}
\Delta J &=& \frac{27}{4\pi}\left(\frac{\sqrt{3}}{4}+\epsilon\right)M\delta M -\frac{9\sqrt{3}}{32\pi}\left(M^{2}\,\epsilon^2+\delta M^2\right) \, .\nonumber\\
\end{eqnarray}}
This clearly shows that the first term dominates over the second term. Thus $\Delta J > 0$ always, indicating that there exists a parameter window allowing particles with appropriate parameters to fall in to overspin it.

Thus a six dimensional pure GB black hole could indeed be overspun under linear order test particle accretion and thereby violating WCCC. This is generally the case including the Kerr black hole in four dimension that it is possible to violate WCCC at the linear order which is always restored at the non-linear order accretion. That's the question we take up next.

\subsection{Linear and non-linear order particle accretions through the new version of gedanken experiment}
\label{Sec:non-linear}

Following the remarkable new gedanken experiment developed by Sorce and Wald \cite{Sorce-Wald17} we consider the first and second order particle accretions in order to obtain the correct and true result, whether pure GB rotating black hole in $D=6$ could be overspun or not. According to this work, black hole always favours no overspinning when non-linear perturbations are included, and hence the WCCC is always respected at the non-linear order. Thus, it is of interest to understand how pure GB rotating black hole responds to non-linear order corrections? As expected the result would turn out to positive;i.e. no overspinning is allowed. 

As before we start with a nearly extremal black hole and redefine Eq.~(\ref{Eq:extr-ty}) as a function of black hole mass $M$ and angular momentum $J$
\begin{eqnarray}\label{Eq:f}
f=243 M^4-1024 J^2\pi^2 \, ,
\end{eqnarray}
and $f\geq 0$ defines black hole with equality indicating   extremality. When $f<0$ there exists no horizon leading to naked singularity while $f=0$ corresponds to the extremality condition. 

Following the Sorce and Wald~\cite{Sorce-Wald17}, we begin with the one-parameter family of perturbation function $f(\lambda)$ including linear and non-linear perturbations. The one-parameter family function allows for deviation from the initial value of $f$, and hence recalling Eq.~(\ref{Eq:f}) we write 
\begin{eqnarray} \label{Eq:func1}
f(\lambda)&=& 243 M(\lambda)^4-1024J(\lambda)^2\pi^2 \, ,
\end{eqnarray}
where $M(\lambda)$ and $J(\lambda)$ are black hole's final parameters due to falling in of matter and given by
\begin{eqnarray} \label{Eq:func2}
M(\lambda)&=& M+\lambda~\delta M\, ,\nonumber\\
J(\lambda)&=&J+\lambda~\delta J\, .
\end{eqnarray}
Note that $\delta M$ and $\delta J$ are first order particle perturbations. As seen from Eq.~(\ref{Eq:func1}) one can write $f(0)=243M^4\epsilon^2$ for sub-extremal black hole with small $\epsilon$, i.e. $\epsilon \ll 1$. Let us then expand $f(\lambda)$ up to quadratic order in $\epsilon$ and  $\lambda$ to define the first and second order particle perturbations as
 \begin{eqnarray}\label{Eq:f3}
f(\lambda)=243M^4\epsilon^2+f_1\lambda+f_2\lambda^2+O(\lambda^3, \lambda^2\epsilon,\lambda\epsilon^2,\epsilon^3)\, .
\end{eqnarray}
We shall now show that when only linear order perturbations are involved, $f(\lambda)<0$ indicating overspinning leading to naked singularity which gets reversed when non-linear perturbations are included. 
To show that we analyse the function $f(\lambda)$ in terms of linear $f_1$ and non-linear $f_2$ perturbations,
 \begin{eqnarray}\label{Eq:liner-per}
f(\lambda)&=&243M^4\epsilon^2+2\left[243 M \delta M-1024 \pi ^2 J \delta J \right]\lambda  \nonumber\\&+&\left[243  M\delta ^2 M-1024 \pi ^2  J\delta ^2 J\right.\nonumber\\&+&\left. 243 \delta M^2-1024 \pi ^2 \delta J^2\right]\lambda^2+O(\lambda^3, \lambda^2\epsilon,\lambda\epsilon^2,\epsilon^3)\, .\nonumber\\
\end{eqnarray}
As shown in Ref. \cite{Sorce-Wald17} we consider an optimal choice of linear particle accretion, according to which falling in particle can carry the minimum possible value of $\delta M$ into the black hole. For the optimal choice we get $\delta M  = {a}/{2\,\mathcal{M}^{1/2}r_{+}^{\,3/2}}$ and keep the terms  up to quadratic order in $\epsilon$. For the sake of clarity, we explore $f(\lambda)$ numerically for linear order particle accretion due to the complicated expression of $\delta M$. To this aim we choose $\delta J=J\epsilon$ which is consistent with the test particle approximation. In doing so, we set $M=1$ and have
\begin{widetext}
\begin{eqnarray}\label{Eq:linear1}
f(\lambda)&=& 243 \epsilon^2+ \left[27 J^2  \left(72 \left(\frac{\left(\sqrt{177} B+59 \left(\sqrt{6} C+9\right)\right)^2}{501264 \pi ^2}+4 J^2\right)-\frac{D \left(\sqrt{\frac{3}{59}} B+\sqrt{6} C+9\right)\epsilon  }{\pi ^2}\right)\right.\nonumber\\&\times &\left.  \left(\frac{\Big(\sqrt{177} B+59 \left(\sqrt{6} C+9\right)\Big)^2}{250632\,\pi ^2}+8 J^2\right)^{-2}-2048 \pi ^2  J^2\right]\epsilon\lambda + O(\lambda^3, \lambda^2\epsilon,\lambda\epsilon^2,\epsilon^3)\, ,
\end{eqnarray}
\end{widetext}
where
\begin{widetext}
\begin{eqnarray}
B&=& \Big[1593+512\ 59^{2/3} \pi ^{8/3}+472 \left(\sqrt[3]{59} \pi ^{4/3}-8 \pi ^2\right) J^2\Big]^{1/2}\, ,\\
C&=& \Bigg[\frac{9 \sqrt{177} \left(9-32 \pi ^2 J^2\right)}{B}+\frac{32 \sqrt[3]{59} \pi ^{8/3} \Big(216 \sqrt[3]{59}-\left(\sqrt[3]{59}+8 \pi ^{2/3}\right) J^2 \left(59 J^2+512 \pi ^2\right)\Big)}{B^2}\nonumber\\&+&\frac{1593 \Big(4 \left(\sqrt[3]{59} \pi ^{4/3}-32 \pi ^2\right) J^2+27\Big)}{B^2}\Bigg]^{1/2}\, , \\
D&=& \pi ^{4/3} \left[\frac{59 \sqrt{2} C \left(64 \sqrt[3]{59} \pi ^{4/3}-59 J^2\right) \Big(531 \sqrt{177} \left(32 \pi ^2 J^2-9\right)-B^3\Big)}{B \left(236 \left(\sqrt[3]{59} \pi ^{4/3}+16 \pi ^2\right) J^2+256\ 59^{2/3} \pi ^{8/3}-1593\right)+531 \sqrt{177} \left(32 \pi ^2 J^2-9\right)}\right.\nonumber\\&+&\left. 2 \sqrt{59} B \left(59 J^2-64 \sqrt[3]{59} \pi ^{4/3}\right)\right]\, .
\end{eqnarray}
\end{widetext}

Since near extremality is defined by $f(0)=243M^4\epsilon^2$ from Eq.~({\ref{Eq:func1}}), and hence angular momentum will have the form as
\begin{eqnarray}\label{Eq:num1}
J\text{=}\frac{9 \sqrt{3} \sqrt{M^4-M^4 \epsilon ^2}}{32 \pi }\, .
\end{eqnarray}
Let us now evaluate  Eq.~(\ref{Eq:num1}) numerically. For given $\delta J=J\epsilon$ with $\epsilon=0.01$, we have $J=0.155053$. Then we evaluate (\ref{Eq:linear1}) and it turns out to be $f(\lambda)=-0.0239842$; thus, we are able to reach $f(\lambda)\leq 0$ for linear order perturbation in $\lambda$.  With this one can conclude that black hole can be overpsun under linear order.

Let us next consider the non-linear order particle accretion including perturbations up to the second order in $\lambda$. Note that we took into account null energy condition, $\delta M-\Omega \delta J = \int_{H} \Xi_{\alpha} \chi_{\beta} \delta T^{\alpha \beta}\geq 0$ with volume element at the horizon $\Xi_{\alpha}$, for linear order perturbation to satisfy $\delta M\geq\Omega \delta J$. Taking the null energy condition to the second order perturbations, we write the variational inequality $\delta^2 M-\Omega \delta^2 J\geq -\frac{k}{8\pi}\delta^2 A$~\cite{Sorce-Wald17}. The surface gravity $\kappa$ and the horizon area $A$ are respectively given in Eqs.~(\ref{Eq:hor-area}) and (\ref{Eq:sur-gravity}).

Now we recall Eq.~(\ref{Eq:f3}) and for inclusion of non-linear order perturbations, it takes the following form
%
 \begin{eqnarray}\label{Eq:non-linear-aneqb}
 f_2&=& 243  M\left(\delta^2 M-\Omega \delta^2 J\right) + 243 \delta M^2-1024 \pi ^2 \delta J^2 \, .\nonumber\\
\end{eqnarray}
In view of the second order null energy condition,  $\delta^2 M-\Omega \delta^2 J\geq -\frac{k}{8\pi}\delta^2 A$ and Eqs.~(\ref{Eq:hor-area}) and (\ref{Eq:sur-gravity}), we write $f(\lambda)$ as follows:
\begin{widetext}
\begin{eqnarray}\label{Eq:linear-perturbation1}
f(\lambda)&\simeq & 243 \left(\epsilon+ \left[	\left(4J^2  \left(\frac{\left(\sqrt{177} B+59 \left(\sqrt{6} C+9\right)\right)^2}{501264 \pi ^2}+4 J^2\right)-\frac{D \left(\sqrt{\frac{3}{59}} B+\sqrt{6} C+9\right)\epsilon  }{18\pi ^2}\right)\right.\right.\nonumber\\&\times &\left. \left. \left(\frac{\Big(\sqrt{177} B+59 \left(\sqrt{6} C+9\right)\Big)^2}{250632\,\pi ^2}+8 J^2\right)^{-2}-\frac{1024}{243} \pi ^2  J^2\right]\lambda\right)^2 + O(\lambda^3, \lambda^2\epsilon,\lambda\epsilon^2,\epsilon^3)\, ,
\end{eqnarray}
\end{widetext}
This clearly shows that $f(\lambda)\geq 0$ is always satisfied when non-linear order perturbations are included.  Thus, $f(\lambda)\geq 0$ always, thereby indicating that black hole is never able to go past the extremality. Thus WCCC that can be violated at the linear order is fully restored when non-linear order particle perturbations are included.


\section{Discussion and conclusions}
\label{Sec:Conclusions}

Even though there is no exact solution of pure Lovelock vacuum equation for a  rotating black hole, following Refs~\cite{Dadhich13b,Dadhich13GB} we have however written an effective metric in terms of the Myers-Perry solution \cite{Myers-Perry86} by incorporating the pure Lovelock potential, $\mu/r^{(D-2N-1)/N}$ in place of the Einstein potential, $\mu/r^{D-3}$. This has all the desirable features of a rotating black hole and it satisfies the equation in the leading order. We have employed this metric to probe WCCC in the pure Lovelock gravity.

In higher dimensional rotating black hole, the critical new aspect that comes in is the fact that gravitational force beyond a threshold dimension could turn repulsive because of repulsive contribution due to rotation overriding the attractive contribution due to mass. For the Einstein gravity, this dimensional threshold is given by $D>5$ while for the pure Lovelock it is $D>4N+1$ which includes Einstein for $N=1$. It turns out that when this happens, black hole cannot overspin even for the linear order particle accretion. This is because in the repulsive environment particles with suitable parameters are unable to reach the black hole horizon. Recently we have shown this elsewhere by an explicit calculation~\cite{Shaymatov20a} that six dimensional Einstein rotating black hole with two rotations cannot be overspun even in  the linear test particle accretion. It may though be noted that for single rotation, even five dimensional black hole cannot be overspun that however happens for a different reason of black hole having only one horizon.

In higher dimensions for rotating black hole, there are two constraints to be satisfied for overspinning being possible even at linear order accretion, and they are: (a) there must exist two horizons;i.e. from Eq. (8), $2N(n+1)+1 > D$ ($n=[(D-1)/2]$ are the maximum number of rotations in dimension $D$) and (b) gravitational force should remain attractive all through outside black hole horizon; i.e from Eq. (9), $4N+1 > D$. It is easy to see that $2N(n+1)+1 \geq 4N+1$ for $n\geq 1$. For overspinning to be unattainable one of these conditions has to be violated. When (a) is violated, black hole has only one horizon and hence the question of overspinning does not arise. On the other hand, when (b) is violated while (a) holding good yet overspinning is not permitted because of particles with proper rotation parameters cannot reach the horizon because of resultant gravity being repulsive. This is what is the case for six dimensional rotating black hole with two rotations in the Einstein gravity. In the case of five dimensional black hole with single rotation, $n=1$,  it is (a) that is violated which prohibits overspinning of the black hole.  

For the pure Lovelock case there is an allowed dimensional window,  that ranges from $D=2N+2$ to $D=4N+1$ as given in Eq.~(\ref{Eq:Dw}), for black hole to be overspun in linear order accretion\footnote{For $N=1$, this window includes $D=4,5$ while for $N=2$, it is $D=6,7,8,9$, and so on for higher orders.}. Note that we have shown by explicit calculation that a six dimensional pure GB rotating black hole could indeed be overspun at the linear order which is however overturned when second order perturbations are included. It is interesting to note that a six dimensional rotating black hole in the Einstein gravity cannot be overspun while it could be in the linear order for the pure GB gravity. This happens simply because the constraint (b) is violated for the former but not for the latter.

Whenever $D>4N+1$, the resultant gravity is repulsive and that is why black hole cannot be overspun because particles that could tend to overspin the black hole would not be able to reach the horizon. This would be true for all $D>4N+1$, and thus WCCC would be always obeyed in all dimensions greater than $4N+1$. This however raises a further much deeper question, how are rotating black holes in $D>4N+1$ formed? They cannot obviously be formed from gravitational collapse because of overall gravitational repulsion. Let's close by posing an important question, rotating black holes may not perhaps exist in dimensions greater than $4N+1$?

\section*{Acknowledgments}
The research of S.S. is partially supported by the Grant No. F-FA-2021-432 of the Uzbekistan Ministry for Innovative Development. N.D. wishes to acknowledge the support of CAS President's International Fellowship Initiative Grant No. 2020VMA0014, and would also like to thank Albert Einstein Institute, Potsdam-Golm for a summer visit.

\bibliographystyle{apsrev4-1}  
\bibliography{gravreferences}

 \end{document}